\documentclass[12pt]{article} 
\usepackage{epsfig}

\newcounter{multieqs}

\newcommand{\bq}{\begin{equation}}
\newcommand{\fq}{\end{equation}}
\newcommand{\bqr}{\begin{eqnarray}}
\newcommand{\fqr}{\end{eqnarray}} \newcommand{\non}{\nonumber
\\} \newcommand{\noi}{\noindent}

\newcommand{\bra}[1]{\langle #1|} \newcommand{\ket}[1]{|#1
\rangle} 
\newcommand{\xpv}[1]{\langle #1 \rangle}

\newcommand{\rf}[1]{(\ref{#1})}

\def\npb#1#2#3{Nucl. Phys. {\bf{B#1}} (#2) #3}

\def\prd#1#2#3{Phys. Rev. {D \bf{#1}} (#2) #3}

\def\ijmp#1#2#3{Int. J. Mod. Phys. {\bf{A#1}} (#2) #3}

 \def\bet{\beta} 
\def\del{\delta} \def\eps{\epsilon} 
 \def\th{\theta} 
  
 \def\sig{\sigma}

\def\Gam{\Gamma} \def\Del{\Delta}

\def\cJ{{\cal J}}  
  
  \def\cR{{\cal R}}

\def\pa{\partial}

\def\rar{\rightarrow} 

\newcommand{\tr}{\mbox{Tr}}

\def\hlf{\frac{1}{2}} \def\ove#1{\frac{1}{#1}}

\begin{document}

\thispagestyle{empty}

\marginparwidth = .5in


\begin{flushright}
\begin{tabular}{l}
ITP-SB-98-57 \\ hep-th/9810115
\end{tabular}
\end{flushright}

\vspace{18mm}
\begin{center}

{\Large \bf Trace anomalies and the string-inspired definition of
quantum-mechanical path integrals in curved space.}

\vspace{18mm}

{K. Schalm}\footnote{E-mail address:
konrad@insti.physics.sunysb.edu} and {P. van
Nieuwenhuizen}\footnote{E-mail address:
vannieu@insti.physics.sunysb.edu} \\[10mm] {\em Institute for
Theoretical Physics \\ State University of New York \\ Stony
Brook, NY 11794-3840, USA }

\vspace{20mm}

{\bf Abstract}
\end{center}

\noi We consider quantum-mechanical path integrals for non-linear
sigma models on a circle defined by the string-inspired method of
Strassler, where one considers periodic quantum fluctuations about a 
center-of-mass
coordinate. In this approach one finds incorrect answers for the local 
trace anomalies of the
corresponding $n$-dimensional field theories in curved space. The quantum field
theory approach to the quantum-mechanical path-integral,
where quantum fluctuations are not periodic but vanish at the endpoints,
yields the
correct answers. We explain these results by a detailed analysis of general
coordinate
invariance in both methods. Both approaches can be derived from the same operator expression and the integrated trace anomalies in both schemes
agree. In the
string-inspired method the integrands are not invariant under general
coordinate transformations and one is therefore not permitted to use Riemann normal coordinates.

\vfill

\newpage
\setcounter{footnote}{0}
\section{Introduction}

One-dimensional (quantum-mechanical) path integrals in curved
space (non-linear sigma models) have been used to compute the chiral
\cite{alg} and trace \cite{bas1} anomalies of $n$-dimensional
quantum field theories coupled to external gravitation and Yang-Mills
fields. In
addition, quantum-mechanical path integrals in flat space have proven useful to
calculate one- and higher-loop Green's functions for field theories
\cite{stras, schub}. In both cases one considers the evaluation of
the partition function of the one-dimensional theory on a
finite time interval $[-\bet_1,\bet_2]$.  

For the evaluation of anomalies
the quantum-mechanical path integral has been defined, in analogy to
quantum field
theory (QFT), by time slicing and requiring that all paths attain the same
values at
$-\bet_1$ and
$\bet_2$ for bosonic point particles (or opposite values for fermionic point
particles)  \cite{deboer}.\footnote{In the case of chiral anomalies, the
fermionic point
particles attain the same values at $- \beta_1$, and $\beta_2$.}  One
decomposes the
paths (``fields")
$x^i(t)$ into a background solution of the free theory satisfying the boundary
conditions at
$-\bet_1$ and $\bet_2$ and quantum deviations vanishing at these points.
Compactifying
the interval to a circle (as seems natural for evaluating the trace, which
yields the
partition function), such paths will in general possess a
kink-discontinuity because
their derivatives need not be continuous at the point where the endpoints come
together. The kinetic operator in the space of functions which
vanish at the endpoints has no zero modes, and is
readily inverted. To compute the partition function one
finally integrates over all possible boundary values.

On the other hand, for the evaluation of Green's
functions in flat space, a string-inspired approach to the path
integral has been used \cite{stras,schub}.  Here the
periodicity of the partition function is reflected in the fact
that bosonic paths are decomposed into periodic fluctuations about a
constant center-of-mass mode (and fermionic paths are expanded into
antiperiodic
fluctuations).\footnotemark[1] The propagator for these deviations is constructed by
inverting the
kinetic operator in the space of periodic functions orthogonal to the constant
modes, which are zero modes of the one-dimensional Laplacian,
and only at the very last does one integrate over the center-of-mass
coordinate.

In this letter we consider quantum-mechanical path integrals
in curved space using the string-inspired definition of path
integrals; we will use the calculation of trace anomalies to
illustrate our points. Both definitions can be derived from the same operator expression by time-slicing. At first sight one would thus expect that
this method should give the same results as the QFT approach. In fact it has
been observed that for linear sigma models (flat-space) the target space
Green's functions in both methods differ by total derivatives
\cite{schub} and we have earlier shown why chiral anomalies
are independent of the approach one uses
\cite{us}. We find that the results for local trace anomalies
differ, due to a subtlety having to do with general coordinate
invariance (Einstein invariance) in curved space.
Namely, in the QFT approach the Riemann summands
$\sqrt{g(x_0)}\bra{x_0}\exp(-\frac{\bet}{\hbar}\hat{H})\ket{x_0} d^n x_0$
of the (discretized) partition function $Z(\bet) = \tr\,
exp(-\frac{\bet}{\hbar}\hat{H})$ are Einstein
invariant, whereas in the string-inspired approach the
corresponding expressions are not Einstein invariant (even if $\hat{H}$ is
Einstein invariant, i.e., even if $\hat{H}$ commutes
with the generator of general coordinate
transformations).  Consequently, one cannot simplify the
calculations in the string-inspired approach by using normal
coordinates.  Only the 
integrated expressions in this approach are general
coordinate invariant.  In the QFT approach, on the
other hand, one obtains directly the correct local trace
anomaly, and the invariance of the summands allows normal
coordinates to be used.  Since both approaches are derived from the same expression, this indicates
that again the integrands computed with either method should
differ by a total derivative.  We present explicit two-loop and three-loop
calculations which support our arguments, and our conclusion is
that the easiest way to compute quantum-mechanical
amplitudes in a {\em curved} space-time background is to follow the
QFT approach rather than the string-inspired method.

\section{Two approaches to the path integral}

The basic object to compute is the partition function: the
trace of the transition element
$\bra{z}\exp(-\frac{\beta}{\hbar}\hat{H}) \ket{y}$.
Here $\hat{H}(\hat{x}^i, \hat{p}_i)$ is an
arbitrary but fixed Hamiltonian operator with
$i=1,\ldots,n$. For definiteness we shall choose an Einstein
invariant $\hat{H}$, which means that $\hat{H}$ commutes with
the generator $\hat{G}(\xi^i(\hat{x}))$ for general coordinate
transformations with parameters $\xi^i(\hat{x})$ \cite{witt}. The
Hamiltonian we consider is given by 
\bq 
\hat{H} =
\frac{1}{2}g^{-1/4}(\hat{x})\hat{p}_i
g^{ij}(\hat{x})g^{1/2}(\hat{x}) \hat{p}_j g^{-1/4}(\hat{x})
;~~g(\hat{x})=\det g_{ij}(\hat{x}) \ .
\label{ham}
\fq
In the transition element one inserts $N-1$ complete sets
of $\hat{x}$-eigenstates and $N$-sets of
$\hat{p}$-eigenstates, rewrites the Hamiltonian in
Weyl-ordered form\footnote{It has been shown in \cite{deboer2} that also 
in curved space one may
replace the Weyl-ordered exponential by the exponential of the
Weyl-ordered Hamiltonian, because the difference vanishes for
$N \rar \infty$. This is due to the fact that the propagators
$\xpv{p_i(t_1)p_j(t_2)}$ and $\xpv{p_i(t_1)q^j(t_2)}$ are not
singular in $\eps \equiv \bet/N$. With
$\hat{H}$ in Weyl-ordered form one may use the
midpoint rule to evaluate arbitrary functions of the operators
$\hat{x}$ and $\hat{p}$ in terms of the eigenvalues $x$ and $p$. Weyl-ordering
of~\rf{ham} leads to well-known extra noncovariant terms of order
$\hbar^2$ which are crucial for the general coordinate
invariance of the transition element.}, replaces the $\hat{x}$ and
$\hat{p}$-operators by their $c$-number eigenvalues and
integrates over all $p$'s. The factors
$det(g_{ij}(x_k+x_{k-1}/2))$ produced by the integration over
the $p$'s are exponentiated by anticommuting Lee-Yang ghosts
$b^i,~c^i$ and a commuting Lee-Yang ghost $a^i$ \cite{bas1}.  Since we are
interested in
taking traces, we set $z=y$.

From this point on the QFT approach and the string-inspired
approach go separate ways. In the QFT approach one expands the
$z^i = x_0^i,x_1^i,\ldots,x_{N-1}^i,x_N^i = y^i$ into a constant
background part $x_0^i=z^i=y^i$ satisfying the free field
equations of the free field part of the action
$\frac{1}{2}g_{ij}(x_0)\frac{1}{\eps}(x_k^i-x_{k-1}^i)\frac{1}{\eps}(x_k^j-x_{k-
1}^j)
\sim \frac{1}{2}g_{ij}(x_0)\dot{x}^i\dot{x}^j$ (where $\eps = \bet/N$), and quantum deviations
$q_k^i$ which vanish at
the endpoints $k=0$ and $k=N$. The $q_k^i$ are expanded into
eigenfunctions  that satisfy these boundary conditions
\bq 
q_k^i = \sum_{m=1}^{N-1}r_m^i
\sqrt{\frac{2}{N}}
\sin\left(\frac{km\pi}{N}\right)~;~~k=1,...,N-1~;~i=1,...,n \ .  
\fq
By coupling $\frac{1}{2}(q_k^i+q_{k-1}^i)$ and
$(q_k^i-q_{k-1}^i)/\eps$  to external
sources, one obtains the discretized propagators in closed
form, and by taking the limit $N\rar\infty$ one reads off the
correct Feynman rules \cite{deboer} (the rules how to evaluate
integrals over products of distributions $\th(\sig-\tau)$ and
$\del(\sig-\tau)$ as they appear in Feynman graphs). The
upshot is that $\del(\sig-\tau)$ is to be considered a
Kronecker delta function and that $\th(0)=1/2$, so that for
example (we define $t=\bet\tau$ for convenience) 
\bq
\int_{-1}^0 d\sig\int_{-1}^0 d\tau
\del(\sig-\tau)\th(\sig-\tau)\th(\sig-\tau) = \frac{1}{4} =
\int_{-1}^0 d\sig \int_{-1}^0 d\tau
\del(\sig-\tau)\th(\sig-\tau)\th(\tau-\sig) \ .
\label{ident}
\fq

After a careful detailed analysis \cite{deboer} one finds that 
\bq 
\tr
\bra{z}\exp(-\frac{\beta}{\hbar}\hat{H}) \ket{y} = \int d^nx_0
\sqrt{g(x_0)}
\frac{1}{(2\pi\beta\hbar)^{n/2}}\xpv{e^{-\frac{1}{\hbar}S_{int}}}_{x_0}
\ ,
\label{trans}
\fq 
where\footnote{Our conventions for the Riemann and Ricci
tensor are that $R_{ijk}^{~~~l} =\pa_i\Gam_{jk}^{~~l} +
\Gam_{im}^{~~l}\Gam_{jk}^{~~m} - (i \leftrightarrow j)$ and
$R_{ik} =R_{ijk}^{~~~j}$.}  
\bqr 
-\frac{1}{\hbar}S_{int} &=&
-\frac{1}{2\beta\hbar} \int_{-1}^0
d\tau\left[g_{ij}(x_0+q)-g_{ij}(x_0)\right](\dot{q}^i\dot{q}^j
+ b^ic^j + a^ia^j) \non && -\frac{\beta\hbar}{8}\int_{-1}^0
d\tau \left[R(x_0+q)+ g^{ij} \Gam_{im}^{~~n}\Gam_{jn}^{~~m}
(x_0+q)\right] \ .
\label{sint}
\fqr
The integral $\int d^n x_0 \sqrt{g(x_0)}$ on the r.h.s. of
\rf{trans} comes from taking the trace and the factor
$(2\pi\bet\hbar)^{-n/2}$ coincides with the Feynman measure.
The terms with $R$ and $\Gam\Gam$ are created by Weyl ordering eq.~\rf{ham}.
The continuum limits of
the propagators to be used for the perturbative evaluation of the last
factor are
\bqr
\xpv{q^i(\sig)q^j(\tau)}_{x_0} &=& -\beta\hbar g^{ij}(x_0)
\Del^{QFT} (\sig,\tau) \ ,\non 
\xpv{b^i(\sig)c^j(\tau)}_{x_0} &=&
-2\beta\hbar g^{ij}(x_0) \del(\sig-\tau) = -2
\xpv{a^i(\sig)a^j(\tau)}_{x_0}
\label{prop}
\fqr 
with 
\bqr 
\Del^{QFT}(\sig,\tau) &=&
\sig(\tau+1)\th(\sig-\tau) + \tau(\sig+1)\th(\tau-\sig) \ .
\label{ts-prop}
\fqr
Note that $\xpv{\dot{q}^i(\sig)\dot{q}^j(\tau)} \sim
{}^{\bullet}\Del_{QFT}^{\bullet}(\sig,\tau) =
1-\del(\sig-\tau)$ is singular for $\sig \rar \tau$ but that
this singularity cancels in the sum $\xpv{\dot{q}^i\dot{q}^j}
+ \xpv{b^ic^j} +\xpv{a^ia^j}$. At higher loops the
$a^i,b^i,c^i$ ghosts are crucial to remove all ultraviolet
($\sig-\tau \rar 0$) singularities and as a result all
integrals are finite (as they should be in quantum mechanics,
despite the double derivative interactions in
$\frac{1}{2}g_{ij}(x)\dot{x}^i\dot{x}^j$). Since we work on
the finite interval $[-\bet,0]$ (or $[-1,0]$ for
$\tau=t/\bet$) there are no infrared singularities.

The string-inspired approach to the partition function
starts from the same discretized expression of the transition
element as an integral over the complete sets
$d^nx_1,...,d^nx_{N-1}$, but one then includes the trace
integration over $x_0^i=x^i_N$ and treats all points
$x^i_0,x^i_1,...,x^i_{N-1}$ on equal footing. One thus
considers the same object 
\bq 
\tr
\bra{z}\exp(-\frac{\beta}{\hbar}\hat{H}) \ket{y} = \int d^nx_0
\sqrt{g(x_0)} \bra{x_0}\exp(-\frac{\beta}{\hbar}\hat{H})
\ket{x_0} \ ,
\label{trans2}
\fq
but now one expands the $n \times N$ integration variables $x_k^i$ into
periodic functions which are eigenfunctions of the free action
\bq 
x_k^i =
\sum_{p=0}^{N/2}\frac{2}{\sqrt{N}}\cos\left(\frac{2kp\pi}{N}\right)
r_p^{i,c} + \sum_{p=1}^{N/2-1}
\frac{2}{\sqrt{N}}\sin\left(\frac{2kp\pi}{N}\right)
r_p^{i,s};~~ k =1,...,N \ .
\label{com}
\fq
(We take $N$ to be even.) In the discretized expression for~\rf{trans2}, only
factors of the metric $g_{ij}$ at midpoints occur, and no
explicit $g_{ij}(x_0)$ survive. We decompose $x^i_k$ into a center of mass 
part
$x^i_c = \frac{2}{\sqrt{N}} r^{i,c}_0$ and fluctuations $q^i_k$ around
$x^i_c$.  Coupling
the $q^i_k$ to the same external sources as before, but substituting
then~\rf{com} one
finds after considerable tedious but straightforward algebra the
propagators for the
string-inspired method in closed discretized form \cite{us}. They contain
discretized
theta- and delta-functions with the same properties as before; in
particular~\rf{ident} holds again. The continuum propagators
are now given by~\rf{prop} with 
\bq 
\Del^{SI}(\sig,\tau) =
\frac{1}{2}(\sig-\tau)\eps(\sig-\tau)
-\frac{1}{2}(\sig-\tau)^2 -\frac{1}{12} \ .
\label{com-prop}
\fq
and $x_0$ replaced by $x_c$. As expected, this propagator is translation invariant, and satisfies
$\pa_{\sig}^2
\Del^{SI}(\sig-\tau) = \del(\sig-\tau)-1$ which is the Dirac
delta-function in the space of functions $q(\tau)$ orthogonal
to the constant. The factor $-\frac{1}{12}$ ensures that
$\int_{-1}^0 \Del^{SI}(\sig,\tau) d\tau = 0$, which should be satisfied as
$\langle q
( \sigma ) \int_{-1}^0 q ( \tau ) d \tau \rangle =0$.

The expression for the trace in~\rf{trans2} is then formally
the same as obtained from~\rf{trans}, namely 
\bq 
\int d^nx_0
\sqrt{g(x_0)} \bra{x_0}\exp(-\frac{\beta}{\hbar}\hat{H})
\ket{x_0}= \int d^nx_c \sqrt{g(x_c)}\ove{(2\pi\bet\hbar)^{n/2}}
\xpv{e^{-\frac{1}{\hbar}S_{int}} }_{x_c} \ , 
\label{twoeleven}
\fq
where the right-hand-side is now evaluated using
$\Del^{SI}(\sig,\tau)$. In particular $S_{int}$ is the same
as~\rf{sint} but with $x^i_0$ replaced by $x^i_c$. However, the integrand
$\xpv{\exp(-\frac{1}{\hbar}S_{int}) }_{x_c}$ cannot be written as a
matrix element with well-defined position bras and kets;
rather it is a function which can be computed
using~\rf{com-prop} and~\rf{ident}. In particular the $x_c$
in $\sqrt{g(x_c)}$ is not a true position like the $x_0$ in
$\sqrt{g(x_0})$, which is an eigenvalue of the
$\hat{x}$-operator. It is this fact which is of crucial
importance for the issues of general covariance. Starting with
an Einstein invariant Hamiltonian our
final answer for the partition function will be general
coordinate invariant.  In the QFT case the integrands 
$\langle z | \exp(-\frac{
\beta}{\hbar}\hat{H}) |
y \rangle$ and the tracing measure $\int d^nx_0\sqrt{g(x_0)}$ are 
separately Einstein invariant by construction.  In the string-inspired case 
there is no
reason why
the integrands on the right-hand-side of~\rf{twoeleven} should be Einstein
invariant,
and explicit calculations show that they are not Einstein invariant.  Only
the integrated
expression for the partition function is guaranteed to be covariant 
(and is the
same in both approaches since in both cases one integrates over all $n 
\times N$
variables).

We have now defined the path integral according to the
QFT method and the string-inspired method. Both have
the same non-covariant order $\hbar^2$ counterterm needed for
the Einstein invariance of the final results, and both have
the same vertices because both are based on time-slicing. The
difference rests solely in the order of the integration over
$d^nx_1, d^nx_2,...,d^nx_{N-1}, d^nx_N=d^nx_0$: in the QFT
approach we first integrate over intermediate
$x^i_1,...,x^i_{N-1}$ using~\rf{ts-prop} and then integrate
over $\sqrt{g(x_0)}d^nx_0$, whereas in the string-inspired
method one first integrates over the fluctuations about the
center-of-mass $x^i_c= 1/N\sum x_k^i$ using~\rf{com-prop}, and then over
$\sqrt{g(x_c)}d^n x_c$.  Note that the number of eigenfunctions 
into which quantum
fluctuations are
expanded is the same in both cases (sines and cosines of the double angle
vs. only sines
of the single angle). However, if we view the integration region in both
cases as a
circle, we have a kink in the paths in the first case but not in the second
case.

\section{Comparison of approaches:  Trace anomalies in $n=2$ and $n=4$
dimensions}
\setcounter{equation}{0}

The anomalies of $n$-dimensional quantum field-theories can be
written as 
\bq 
An = \lim_{\bet \rar 0} \tr \cJ
\exp(-\frac{\bet}{\hbar}\hat{\cR}) \ , 
\fq 
where $\cJ$ is the
Jacobian $\pa \del\phi(x)/\pa \phi(y)$ of the fields $\phi (x)$
transforming under the symmetry with $\delta \phi (x)$ and $\hat{H}$ is a
regulator. For consistent anomalies the form of $\hat{\cR}$ is
unique \cite{Diaz} and for local trace (Weyl) anomalies for real scalar
fields one finds that $\hat{\cR}$ is equal to the Hamiltonian~\rf{ham} with 
an improvement term.  The QFT approach then yields \cite{bas1}
\bq 
An_W=\lim_{\bet \rar 0}
\tr\frac{1}{4}(2-n)\sig(\hat{x})\exp(-\frac{\bet}{\hbar}\hat{H})
\label{threethirteen}
\fq
where $\sig(x)$ is the Weyl parameter, normalized to $\del_W g_{ij} =
\sig(x)g_{ij}(x)$
and
\bq 
\hat{H} =
\frac{1}{2}g^{-1/4}(\hat{x})\hat{p}_ig^{1/2}(\hat{x})g^{ij}(\hat{x})\hat{p}_j
g^{-1/4}(\hat{x}) - \frac{1}{2}\hbar^2\xi R(\hat{x}) \ .
\label{ham2}
\fq
The term with $R (\hat{x})$ is the improvement term and
$\xi=\frac{1}{4}(n-2)/(n-1)$. The trace in~\rf{threethirteen}
can be
written as a path integral.  For the QFT approach one finds then that the 
local trace anomaly is proportional to
$\int d^nx_0 \sqrt{g (x_0)} {\sigma (x_0)} \xpv{\exp{(-
\frac{1}{\hbar} S_{int})}}_{x_0}$.

Naively, one might expect a similar expression for the trace
anomaly in the string-inspired method, proportional to $\int d^n x_c
\sqrt{g (x_c)}
\sigma (x_c) \langle \exp {(- \frac{1}{\hbar} S_{int})} \rangle_{x_c} $.  
One should, however, take care in defining the expectation
value of a
 local operator in the string-inspired approach. The naive
 equivalence with the QFT approach by taking the local
 operator at the point $x_c$ is incorrect since at a
 discretized intermediate stage
\bq 
\int
 \frac{1}{N}\sum_{k=1}^N\sig(x_k) \exp(-\frac{1}{\hbar}S) \neq
 \int \sig (x_c =\frac{1}{N}\sum_{k=1}^N x_k)
 \exp(-\frac{1}{\hbar}S) 
\label{bch}
\fq
where $\int$ denotes
 $\prod_{i=1}^n\prod_{k=1}^N\int\sqrt{g(x_k)}dx_k^i$ and $S$
 is the discretized action. Using cyclicity of the trace,~\rf{threethirteen} 
leads to 
the left-hand-side of~\rf{bch}, not to the right-hand-side. Unambigous is the global trace anomaly where $\sig(\hat{x})$ is a constant. In that case the anomaly is just proportional to the partition function and this is the case we shall consider. 

For $n=2$ we need all (connected and disconnected) two-loop
graphs on the worldline.\footnote{The $\beta$ independent terms
in~\rf{trans} come from
$1+ \frac{n}{2}$ loops (one-loop graphs are independent of $\beta$,
two-loop graphs are
proportional to $\beta$, etc.).}  If one uses normal
coordinates, only graphs with the topology of the number 8 and the
counterterm  contribute. The local anomaly is then proportional to
\bqr
\xpv{e^{-
\frac{1}{\hbar} S_{int}}}_{x_0,~n=2} &=& -\frac{\bet\hbar}{6} R(x) \int_{-1}^0 d\tau
[ \Del(\tau,\tau) \left\{ {}^{\bullet} \Del^{\bullet}(\tau,\tau) +
\del(0)
-{}^{\bullet}\Del(\tau,\tau){}^{\bullet}\Del (\tau,\tau)
\right\} ] \non
&& -\frac{\bet\hbar}{8}R(x) \ .  \fqr
The singularities with $\delta (0)$ are well-defined in the discretized
expressions and cancel in both approaches.  Following the QFT approach, and
substituting~\rf{ts-prop} in the above, one obtains the correct answer
$-\ove{12}R$. In the string-inspired method, substituting~\rf{com-prop},
the second
term vanishes due to the fact that
${}^{\bullet}\Del^{SI}(\tau,\tau)=0$ and the final result is
$-\ove{9}R$, which is incorrect.

The results for the calculation in general coordinates are
summarized in figure 1, where we introduce the notation
$\pa^2g =g^{ij}g^{kl}\pa_k\pa_lg_{ij}$, $\pa^i\pa^jg_{ij} =
g^{ik}g^{jl}\pa_k\pa_lg_{ij}$,
$\pa_mg=g^{ij}\pa_mg_{ij},~g_m=g^{in}\pa_ig_{mn}$,
$g^m=g^{mn}g_n$,
$(\pa_ig_{jk})^2=g^{ip}g^{jq}g^{kr}\pa_ig_{jk}\pa_pg_{qr}$ and
similarly for $\pa_ig_{jk}\pa_jg_{ik}$.  For example in the first line
in figure 1 we find the contribution due to expanding $exp \left[
- \frac{1}{2 \bet\hbar}\int_{-1}^0 \frac{1}{2} q^kq^l \pa_k \pa_l g_{ij} (x_0)
 \pa_kg_{ij}(x_0)\left\{\dot{q}^i\dot{q}^j +
b^ic^j+a^ia^j\right\}d\tau\right]$ to first order and taking
the contractions $\xpv{ \dot{q}^i\dot{q}^j + b^ic^j+a^ia^j}$ and $\langle
q^k q^l
\rangle$.
 In the QFT approach one finds
$-\frac{\bet\hbar}{4}\pa^2g(-\frac{1}{6})$ while in the string
approach one obtains
$-\frac{\bet\hbar}{4}\pa^2g(-\frac{1}{12})$. In the last line
one finds the contribution from the order $\hbar^2$ terms in
the action which were produced by Weyl ordering.
\begin{figure}[ht]
$$
\begin{array}{|cllcc|}
\hline \mbox{\underline{Graphs}:} & \mbox{\underline{Tensor
Structure}:} && \underline{QFT}: & \underline{SI}: \\[.15in]
\raisebox{-.2in}{\epsfxsize=0.5in \epsfbox{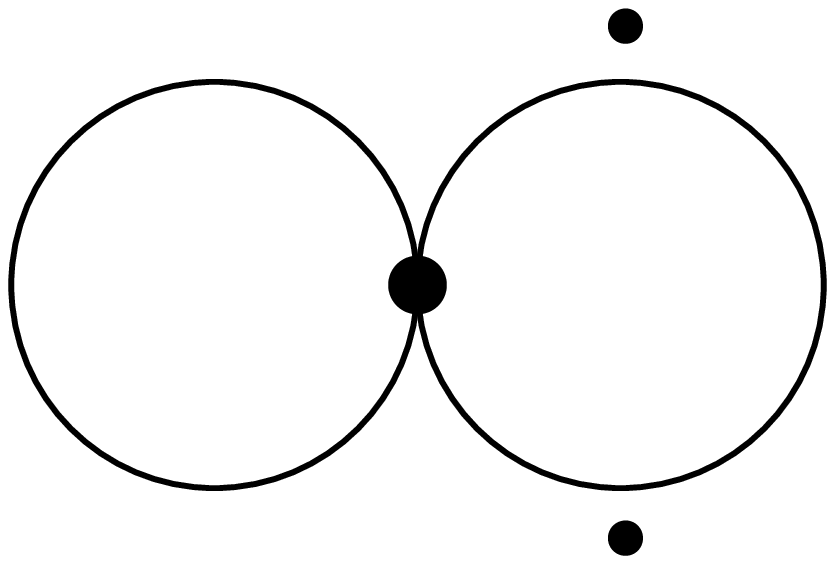}} \;\; +
\;\; \raisebox{-.1in}{\epsfxsize=0.5in \epsfbox{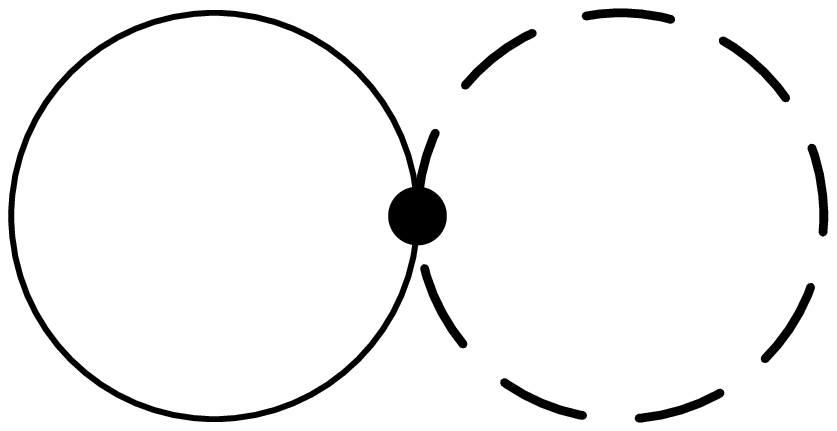}} &
-\frac{\bet\hbar}{4} \pa^2 g && -\frac{1}{6} & -\ove{12}
\\[.15in] \raisebox{-.2in}{\epsfxsize=0.5in
\epsfbox{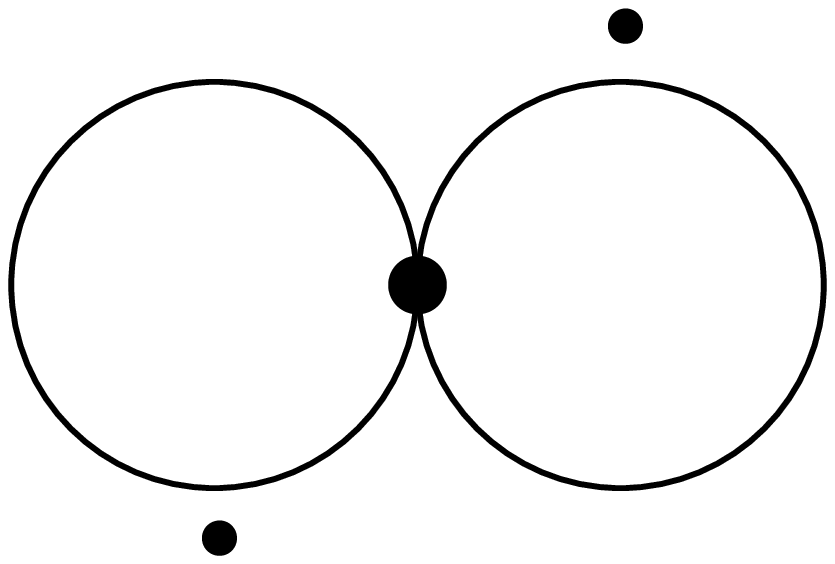}} & -\frac{\bet\hbar}{2}\pa^i\pa^jg_{ij} &&
\frac{1}{12} & 0 \\[.15in]
\left(\raisebox{-.22in}{\epsfxsize=0.5in \epsfbox{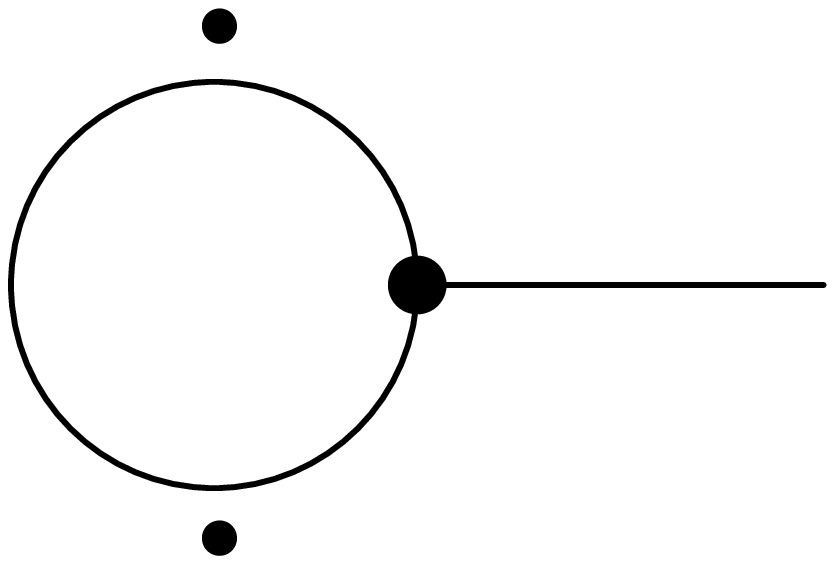}}+
\raisebox{-.1in}{\epsfxsize=0.5in \epsfbox{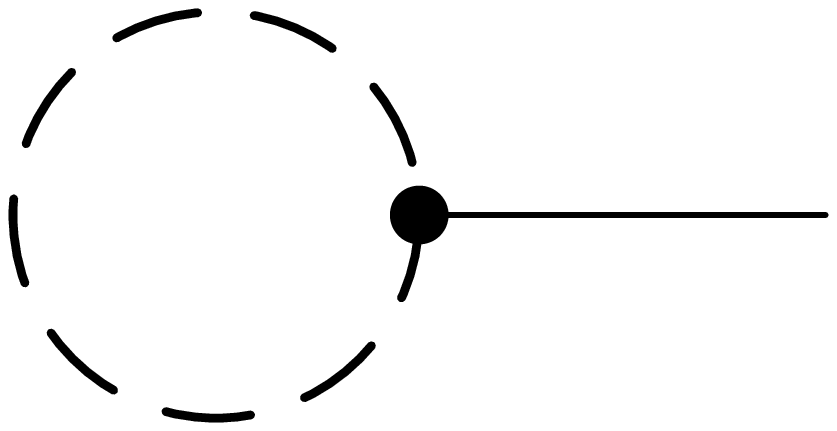}}\right)
\left(\raisebox{-.22in}{\epsfxsize=0.5in \epsfbox{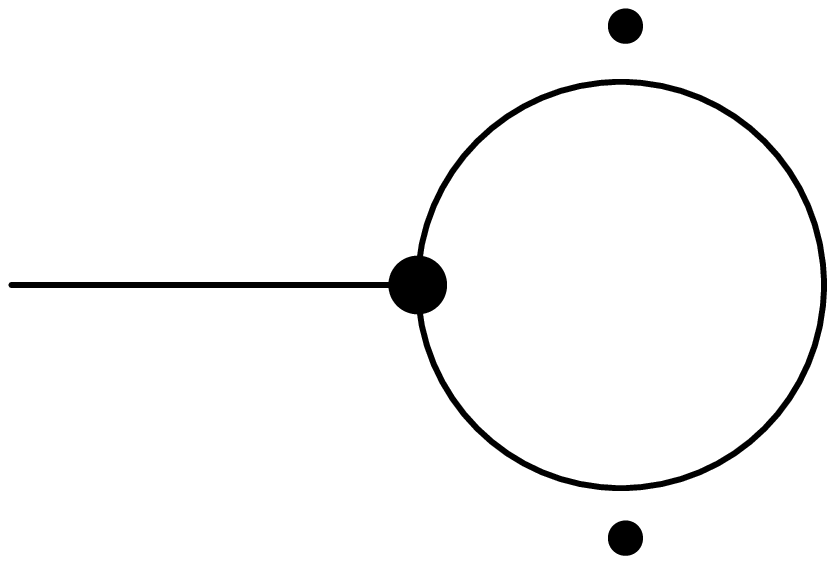}}+
\raisebox{-.1in}{\epsfxsize=0.5in \epsfbox{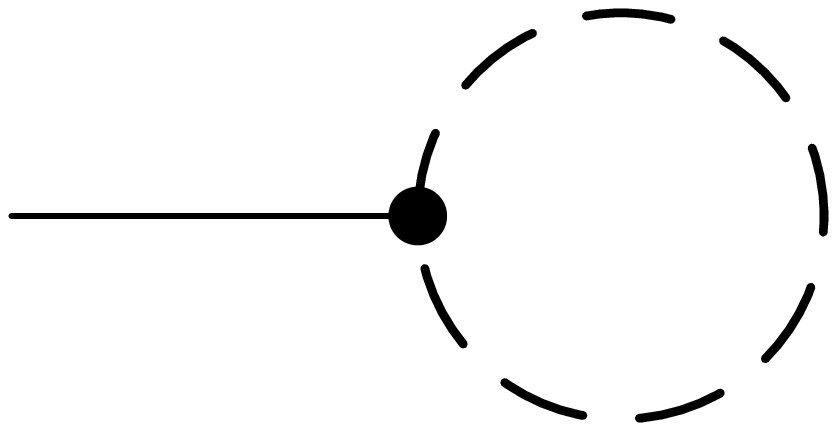}}\right) &
-\frac{\bet\hbar}{8}\pa^mg\pa_mg &&-\frac{1}{12} & 0 \\[.15in]
\left(\raisebox{-.22in}{\epsfxsize=0.5in \epsfbox{mod7.ps}}+
\raisebox{-.1in}{\epsfxsize=0.5in \epsfbox{mod9.ps}}\right)
\left(\raisebox{-.1in}{\epsfxsize=0.5in \epsfbox{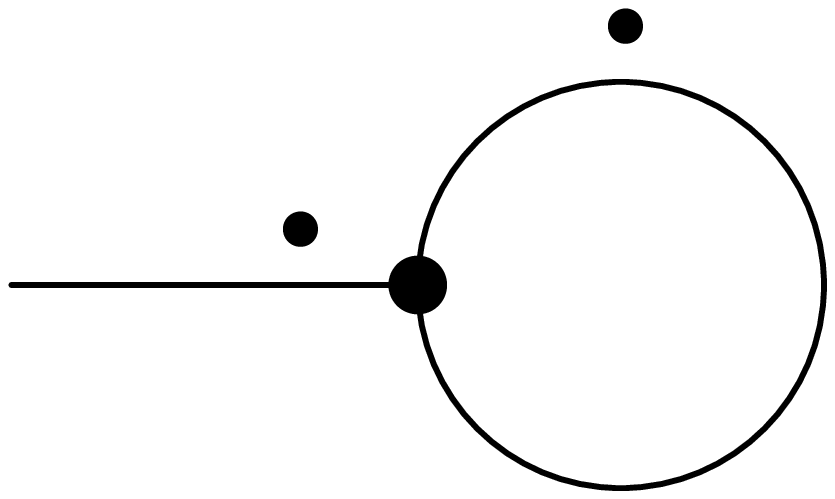}}
\right) & -\frac{\bet\hbar}{2}\pa^mgg_m && \frac{1}{12} & 0
\\[.15in] \raisebox{-.1in}{\epsfysize=0.3in \epsfbox{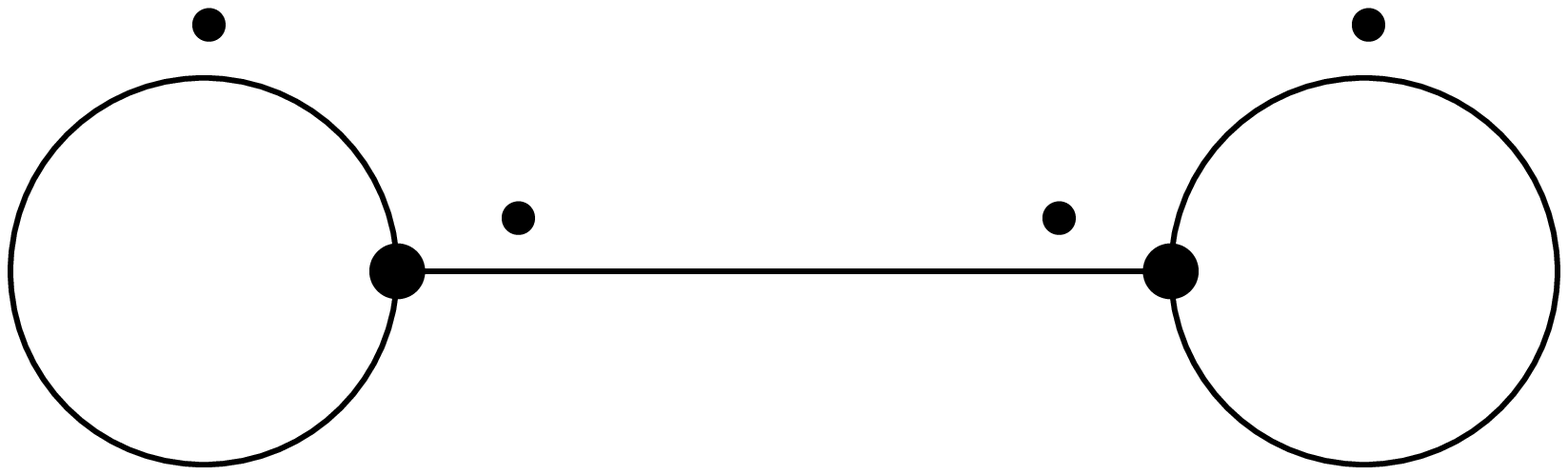}}
& -\frac{\bet\hbar}{2}g^mg_m && -\frac{1}{12} & 0 \\[.15in]
\raisebox{-.1in}{\epsfysize=0.3in \epsfbox{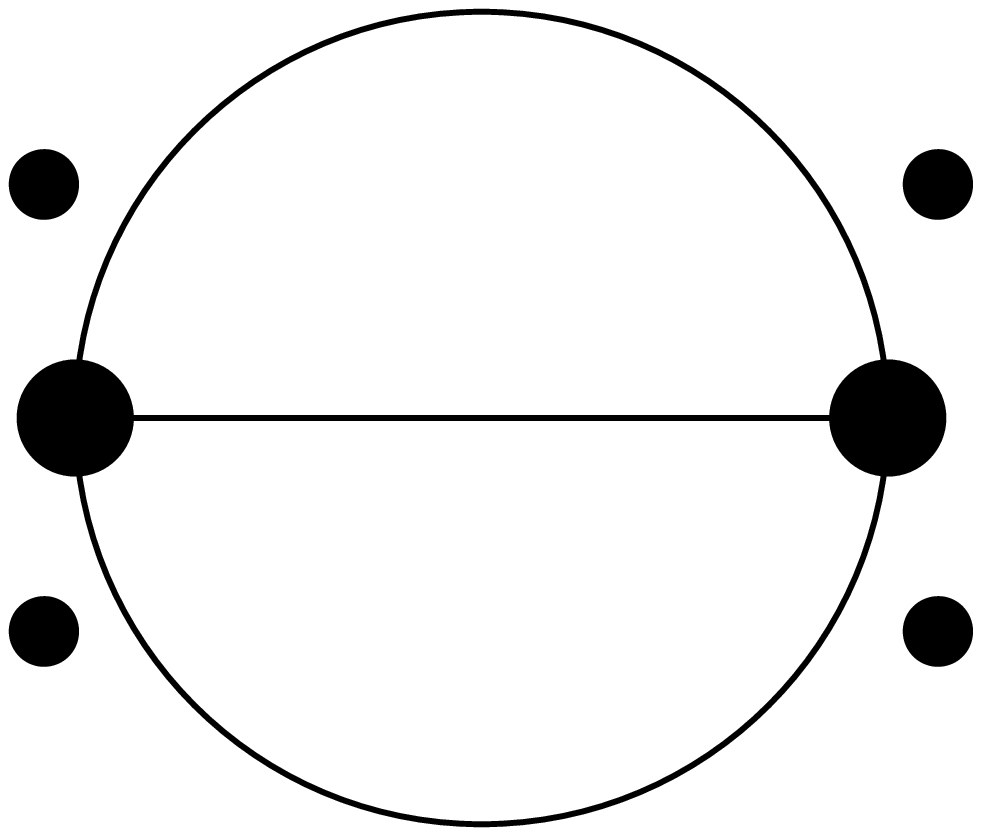}}\;\; -\;\;
\raisebox{-.1in}{\epsfysize=0.3in \epsfbox{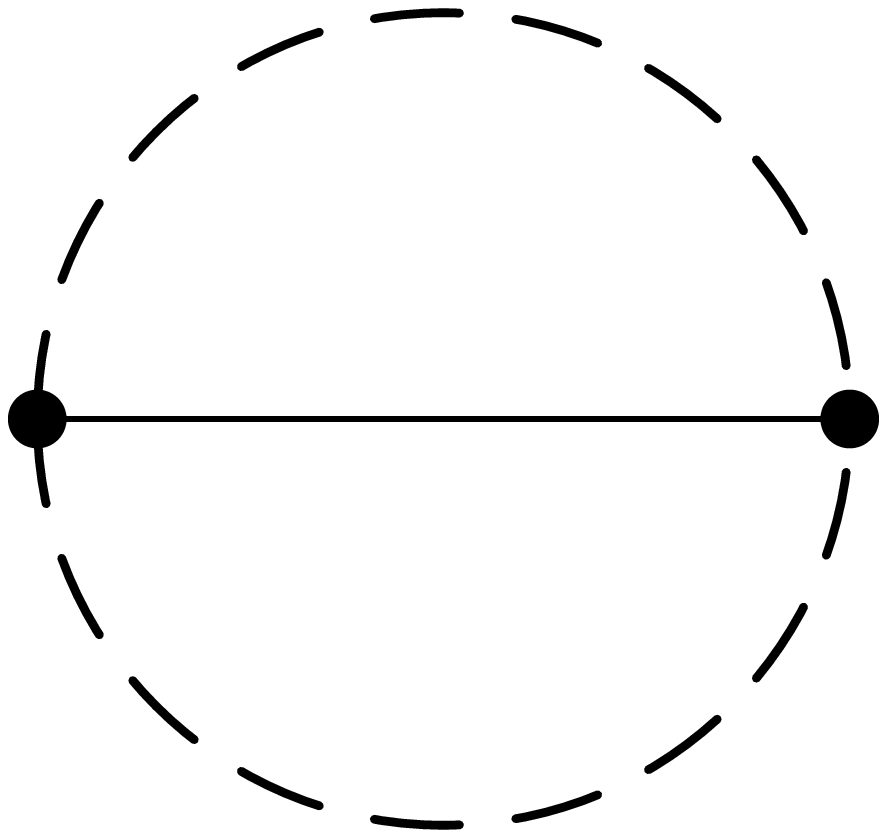}} &
-\frac{\bet\hbar}{4}(\pa_ig_{jk})^2 && \frac{1}{4} & \ove{6}
\\[.15in] \raisebox{-.1in}{\epsfysize=0.3in \epsfbox{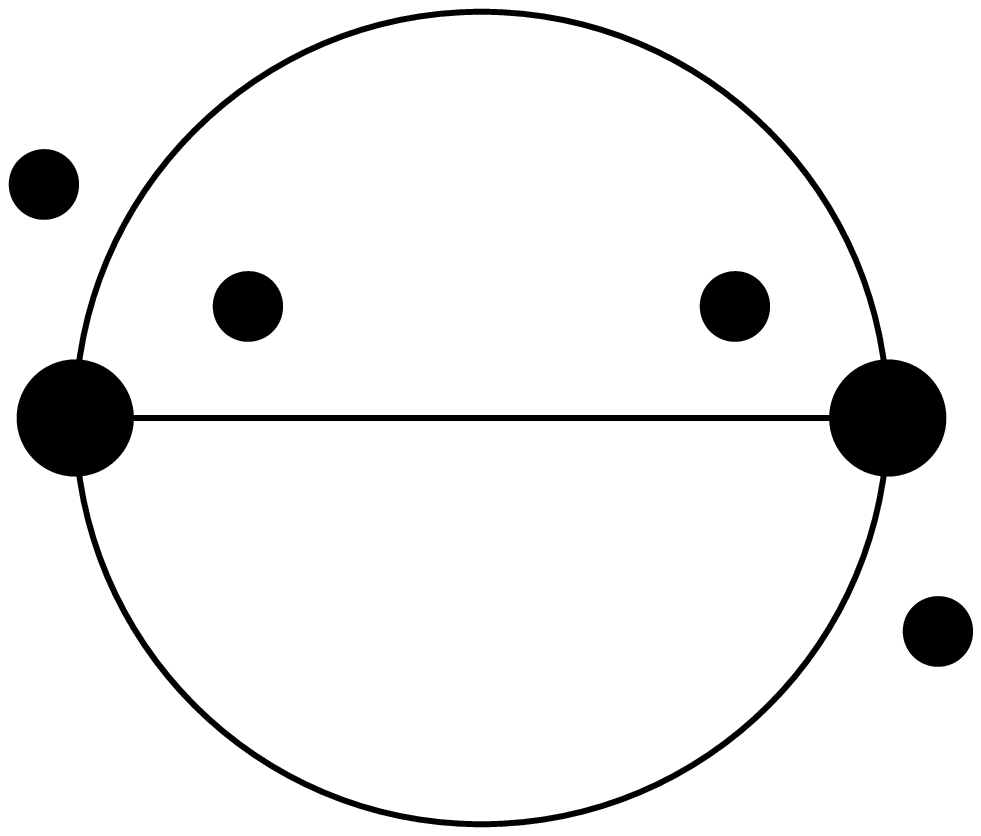}}
& -\frac{\bet\hbar}{2}\pa_ig_{jk}\pa_jg_{ik} && -\frac{1}{6} &
-\ove{12} \\[.15in] \bullet &
-\frac{\bet\hbar}{8}R-\frac{\bet\hbar}{8}g^{ij}\Gam_{ik}^{~~l}\Gam_{jl}^{~~k}
&&1&1 \\[0.1in] \hline
\end{array}
$$
\caption{The results for the trace anomaly for a real scalar
field in $n=2$ dimensions. Dotted lines denote ghosts. For
each graph, the result in either the QFT or string-inspired
(SI) approach is obtained by taking the product of the tensor
structure and the number in the appropriate column. The tensor
structure is the same in both cases as the vertices are the
same. Because $\Del_{SI}^{\bullet}(\tau,\tau)=0$ in the
string-inspired approach, there are fewer nonvanishing
contributions than in the QFT approach. However, the sum of
all contributions is general coordinate invariant in the
QFT but not in the SI approach.}
\end{figure}

Adding all contributions, we find for $\xpv{\exp{(-
\frac{1}{\hbar} S_{int})}}_{x_0}$ 
(proportional to the local Weyl anomaly of a real scalar field in $n=2$ dimensions) in the
QFT approach
an Einstein invariant result, which of course equals that of the normal
coordinate
calculation.
\bqr 
\xpv{e^{-
\frac{1}{\hbar} S_{int}}}_{x_0,~n=2}
&=&
\left(-\frac{1}{4} \pa^2 g\right)\left(-\frac{1}{6}\right)
+\left(-\frac{1}{2}\pa^i\pa^jg_{ij}\right)\left(\frac{1}{12}\right)
\non && +
\left(-\frac{1}{8}\pa^mg\pa_mg\right)\left(-\frac{1}{12}\right)
+ \left(-\frac{1}{2}\pa^mgg_m\right)\left(\frac{1}{12}\right)
\non &&+
\left(-\frac{1}{2}g^mg_m\right)\left(-\frac{1}{12}\right) +
\left(-\frac{1}{4}(\pa_ig_{jk})^2\right)\left(\frac{1}{4}\right)
\non && +
\left(-\frac{1}{2}\pa_ig_{jk}\pa_jg_{ik}\right)\left(-\frac{1}{6}\right)
+
\left(-\frac{1}{8}R-\frac{1}{8}g^{ij}\Gam_{ik}^{~~l}\Gam_{jl}^{~~k}\right)\non
&=& -\frac{1}{12}R \ .
\label{qft-an}
\fqr
(We have written each term as a product of a tensor structure times the
result of
integrations over expressions depending on $\Delta(\sig,\tau)$. Using normal coordinates one need only evaluate the first, second and last graph in figure 1.)  However, in the
string-inspired
approach we find a noncovariant result
\bqr 
\xpv{e^{-\frac{1}{\hbar}S_{int}}}_{x_c,~n=2} &=&
\left(-\frac{1}{4} \pa^2 g\right)\left(-\frac{1}{12}\right) +
\left(-\frac{1}{4}(\pa_ig_{jk})^2\right)\left(\frac{1}{6}\right)
\non && +
\left(-\frac{1}{2}\pa_ig_{jk}\pa_jg_{ik}\right)\left(-\frac{1}{12}\right)
+
\left(-\frac{1}{8}R-\frac{1}{8}g^{ij}\Gam_{ik}^{~~l}\Gam_{jl}^{~~k}\right)
.
\label{si-an}
\fqr
Hence the integrands in both methods differ. Yet
the integrated expressions~\rf{qft-an} and~\rf{si-an} should be the same,
as we explained in the previous section.  By integrating these expressions with
$\int d^nx_0 \sqrt{g(x_0)}$ and $\int d^nx_c \sqrt{g(x_c)}$,
respectively, and using partial integration we may replace terms
with double derivatives of the metric by products of terms
with single derivatives. This yields under the integral sign
the ``equalities'' 
\bqr 
\sqrt{g}\pa^2g & \Leftrightarrow &
\sqrt{g}\left(-\frac{1}{2}\pa^mg\pa_mg +
(\pa_ig_{jk})^2+g^m\pa_mg\right) \ ,\non
\sqrt{g}\pa^i\pa^jg_{ij} & \Leftrightarrow &
\sqrt{g}\left(-\hlf\pa^mgg_m+\pa_ig_{jk}\pa_jg_{ik}+g^mg_m\right)
\ .  \fqr 
One then indeed finds that the
integrated Weyl anomalies agree 
\bq 
\int
dx_0\sqrt{g(x_0)}\xpv{e^{-
\frac{1}{\hbar} S_{int}}}_{x_0,~n=2}=\int
dx_c\sqrt{g(x_c)}\xpv{e^{-\frac{1}{\hbar}S_{int}}}_{x_c,~n=2} \ .  
\fq

The case $n=2$ is rather special as the answer is proportional to 
$\sqrt{g}R^{(n=2)}$ which is itself a total derivative.  Let
us therefore consider the trace anomaly for a real
scalar field in $n=4$ dimensions and compute it using normal
coordinates. This will explicitly demonstrate that one gets
incorrect results from the string-inspired method. As we
explained before, this is due to the fact that one can only
choose normal coordinates at one point, but since the
integrands (functions of $x_c$) are not Einstein scalars, one
cannot use normal coordinates at each point $x_c$.

In $n=4$ the improvement term in~\rf{ham2} is nonvanishing and
effectively converts the term $\frac{\hbar^2}{8}R$ in~\rf{sint} into
$\frac{\hbar^2}{24}R$. The action yields the following
vertices in normal coordinates 
\bqr 
-\ove{\hbar}S_{int} &=&
-\ove{2\bet\hbar}\int_{-1}^0 d\tau
\left[-\ove{3}R_{iklj}q^kq^l-\ove{6}D_mR_{iklj}q^mq^kq^l
\right.\non &&\left. -
\left\{\ove{20}D_mD_nR_{iklj}-\frac{2}{45}R_{iktm}R_{jl~n}^{~~t}\right\}q^mq^nq^k
q^l+\ldots\right]\left(\dot{q}^i\dot{q}^j
+b^ic^j+a^ia^j\right) \non && -\bet\hbar\int_{-1}^0 d\tau
\left(\ove{24}R(x_0+q)+\ove{8}g^{ij}\Gam_{ik}^{~~l}\Gam_{jl}^{~k}(x_0+q)\right)
\ .  
\fqr
The $\bet$-independent term (which yields the trace anomaly) now arises from
three-loop diagrams. Since there are no three-point vertices in normal 
coordinates,
the five-point vertices do not contribute. The four- and
six-point vertices yield the graphs in figure 2. In the
one-but-last line of figure 2 one finds the contribution from
the order $\hbar^2$ vertex
$\ove{24}R(x_0+q)+\ove{8}g^{ij}\Gam_{ik}^{~~l}\Gam_{jl}^{~~k}(x_0+q)$
due to expanding it to second order in $q^i$ and contracting
the two $q^i$'s to an equal time loop. Note that it would have
been incorrect to omit the $\Gam\Gam $ term alltogether by
arguing that it will not contribute in normal coordinates. In
the last line one finds the contribution from the disconnected
graphs; they are proportional to the square of the $n=2$
result, but with the improvement term their contribution
vanishes in the QFT approach. This is coincidental as for the trace 
anomaly in $n=6$ dimensions the contribution from disconnected graphs 
does not cancel. (It is given by the products of lower dimensional trace anomalies, namely $(-\ove{12}R+\ove{10}R)\ove{720}
(R_{mnpq}^2-R_{mn}^2 -\ove{5}D^2R)+\ove{6}(-\ove{12}R+\ove{10}R)^3$ 
where the terms $\ove{10}R$ and $-\ove{5}D^2R$ come from the improvement 
term.)

\begin{figure}[ht]
$$
\begin{array}{|cllcc|}
\hline \mbox{\underline{Graph}:} &&& \underline{QFT}: &
\underline{SI}: \\[.15in] \raisebox{-.1in}{\epsfxsize=1.2in
\epsfbox{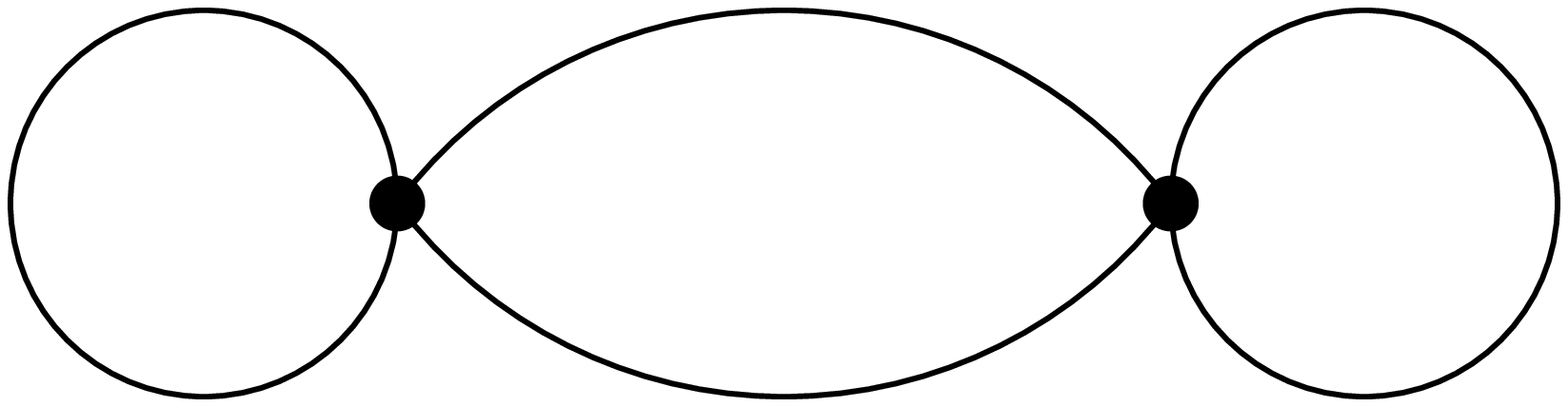}} &&& -\ove{6}R_{mn}^2 &
-\frac{7}{180}R_{mn}^2 \\[0.25in]
\raisebox{-.2in}{\epsfxsize=0.5in \epsfbox{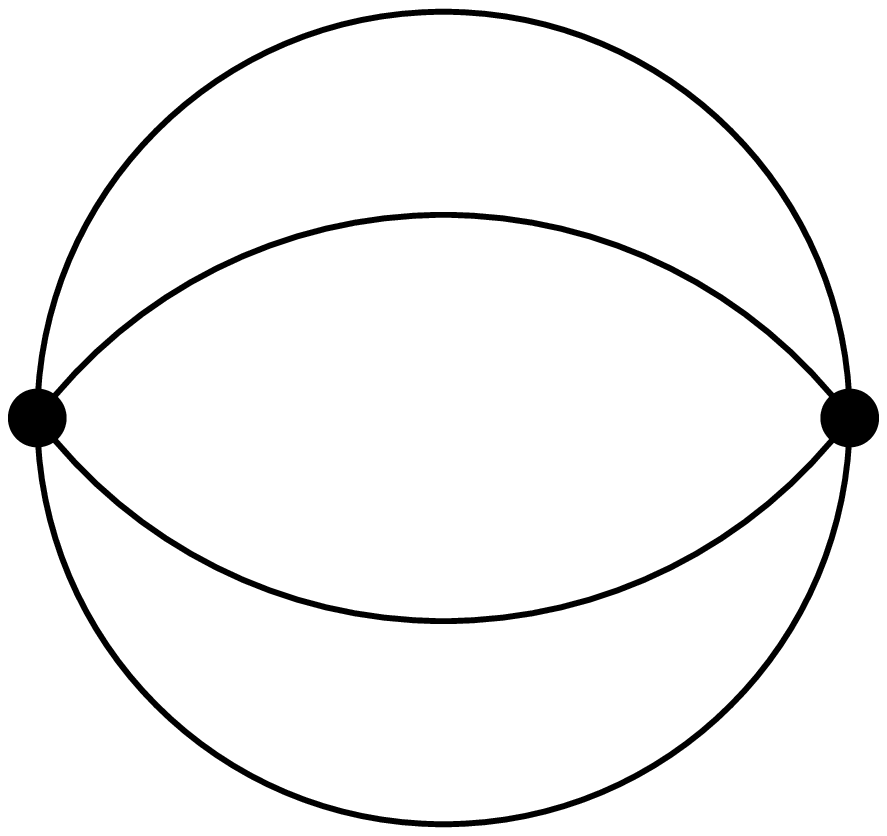}} &&&
-\ove{4}R_{mnpq}^2 & -\ove{60}R_{mnpq}^2 \\[0.25in]
\raisebox{-.25in}{\epsfxsize=0.5in \epsfbox{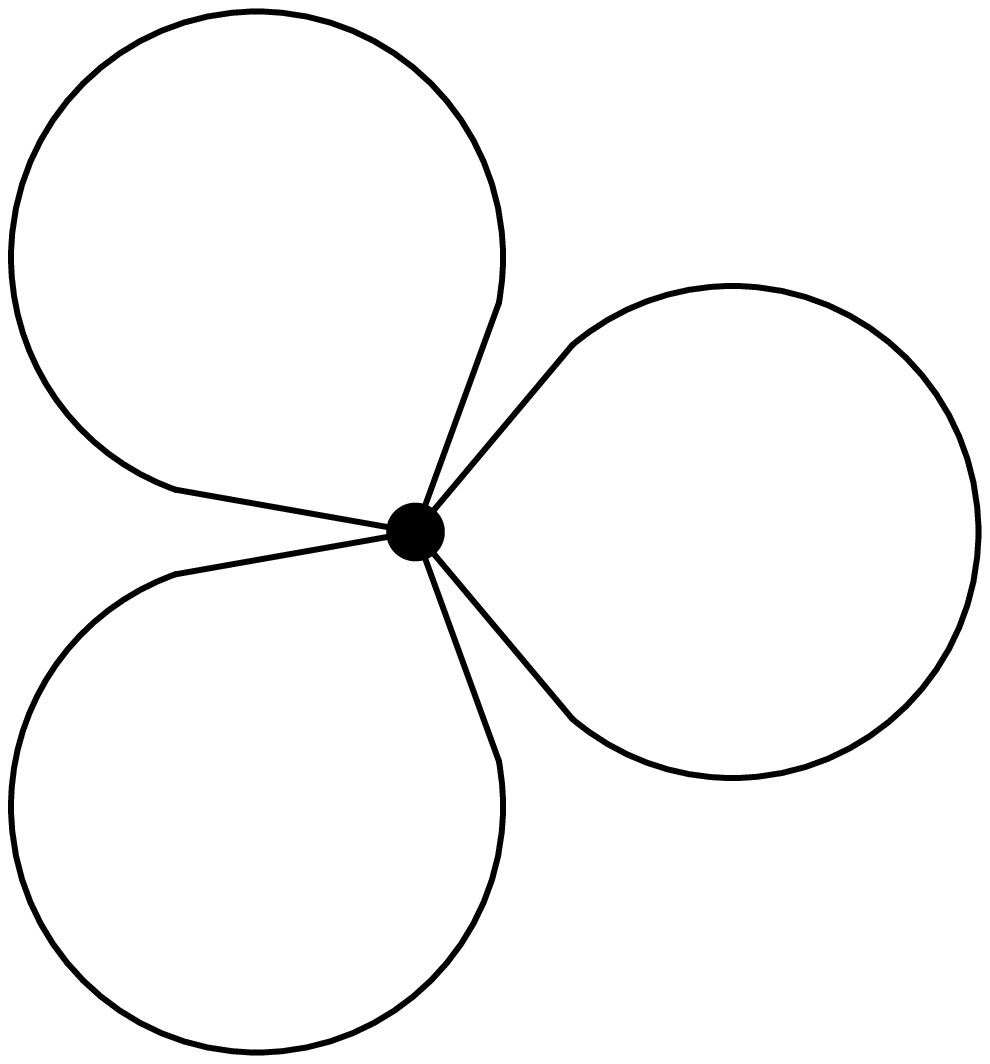}} &&&
\frac{3}{20} D^2R +\ove{10}R_{mnpq}^2 + \ove{15}R_{mn}^2 &
\frac{1}{40} D^2R +\ove{60}R_{mnpq}^2 + \ove{90}R_{mn}^2
\\[.3in] \raisebox{-.15in}{\epsfysize=0.3in
\epsfbox{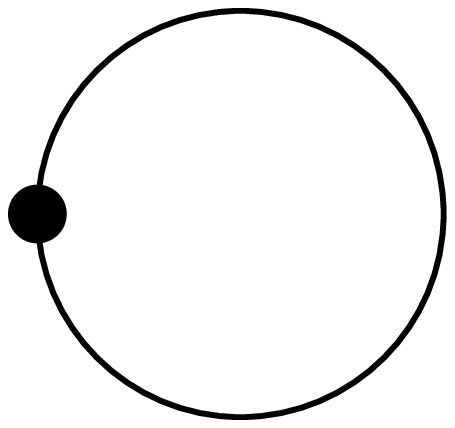}} &&& -\ove{4}D^2R +\ove{4}R_{mnpq}^2 &
-\ove{8}D^2R +\ove{8}R_{mnpq}^2\\[0.25in] \ove{2}
\left(\raisebox{-.1in}{ \epsfysize=0.3in \epsfbox{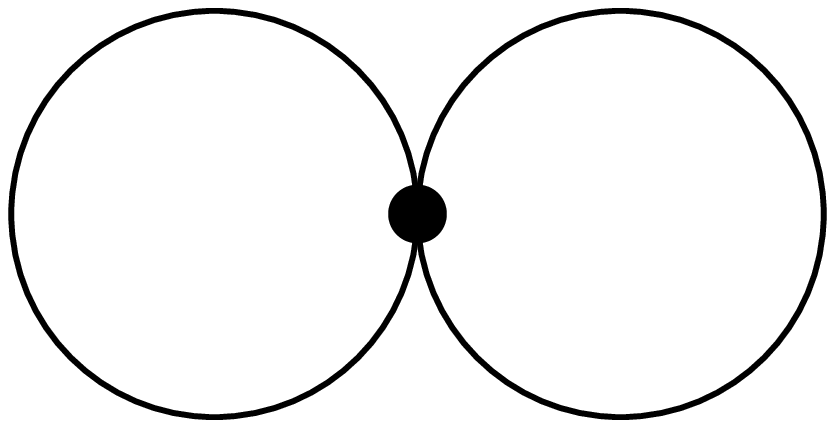}}
\hspace{0.15in} +
\hspace{0.15in} \bullet\right)^2 &&& 0 & \ove{2} \left( \ove{36}R \right)^2
\\[.2in]
\hline
\end{array}
$$
\caption{The results for the trace anomaly for a real scalar
field in $n=4$ dimensions, calculated in Riemann normal
coordinates. Only the topology of the graphs is indicated and
all terms should be multiplied by an overall factor
$(\bet\hbar)^2/72$ to obtain the results in~\rf{qft} and
~\rf{si}.  In the string-inspired approach the fact that
$\Del_{SI}^{\bullet}(\tau,\tau)=0$ again leads to great simplifications but
in this
approach the integrands are not Einstein invariant and the use of normal
coordinates is
therefore illegal.}
\end{figure}

Adding all contributions, we obtain from the QFT approach 
\bqr
\xpv{e^{-
\frac{1}{\hbar} S_{int}}}_{x_0,~n=4}&=&
\frac{(\bet\hbar)^2}{72}\left[\left(-\ove{6}R^2_{mn}\right) +
\left(\frac{3}{20}D^2R+\ove{10}R_{mnpq}^2+\ove{15}R_{mn}^2\right)
\right. \non && \left. + \left(-\ove{4}R_{mnpq}^2\right) +
\left(-\ove{4}D^2R+\ove{4}R_{mnpq}^2\right)\right] \non
&=&\frac{(\bet\hbar)^2}{720}\left[R_{mnpq}^2-R_{mn}^2-D^2R\right]
\label{qft}
\fqr
which is the correct result \cite{bast}. 
In the string inspired
result assuming (incorrectly) that one may use normal coordinates in the
integrands, we
find instead
\bqr 
\xpv{e^{-\frac{1}{\hbar}S_{int}}}_{x_c,~n=4}&=&
\frac{(\bet\hbar)^2}{72}\left[\left(-\frac{7}{180}R^2_{mn}\right)
+
\left(\frac{1}{40}D^2R+\ove{60}R_{mnpq}^2+\ove{90}R_{mn}^2\right)
\right.  \non
&& \left. +\left(-\ove{60}R_{mnpq}^2\right) +
\left(-\ove{8}D^2R+\ove{8}R_{mnpq}^2\right) + \ove{36} R^2
\right] \non
&=&\frac{(\bet\hbar)^2}{72}\left[\frac{1}{8}R_{mnpq}^2-\ove{36}R_{mn}^2-
\ove{10} D^2R + \ove{36} R^2 \right] \ .
\label{si}
\fqr
The $D^2R$ terms come out the same, for which we have no
explanation at hand, but the other terms are different. This
demonstrates that using normal coordinates in the
string-inspired approach yields incorrect results for the
local trace anomaly.

\section{Conclusion}

We have considered path integrals of quantum-mechanical
non-linear sigma models. The operator expression for the partition function 
can be evaluated  in two different
ways: the QFT approach (where paths vanish at the endpoints)
and the string-inspired approach (where paths are manifestly
periodic). At the discretized level, where all expressions are
 well-defined, the difference is that in the QFT approach
one first integrates over the intermediate points $q_k=x_k-x_0$
 for $k=1,...,N$ and then over the endpoint $x_0=x_N$, while in the 
string-inspired approach one first integrates over the deviations 
$q_k=x_k-x_c$ and then over the center-of-mass $x_c=\ove{N}\sum x_k$. 
As the starting point is the same expression for both path integrals they 
should yield the same answers. 

As a test we considered local trace 
anomalies, which are proportional to \linebreak $\tr\, \sig(\hat{x}) 
exp(-\frac{\bet}{\hbar}\hat{\cR})$ with $\hat{\cR}$ a regulator equal to
the Hamiltonian $\hat{H}$ with an improvement term. This yields the correct
 local trace anomalies using the QFT approach. The string-inspired approach 
to path integrals does not reproduce the correct local trace anomaly from 
$\int d^nx_c \sqrt{g(x_c)} \sig(x_c)\xpv{\exp(-\frac{1}{\hbar}S_{int})}_{x_c}$.
We understood this by going back to the discretized level in which case 
$\sig(x_c)$ equals $\sig(\ove{N}\sum x_k)$ and not $\ove{N}\sum\sig(x_k)$. 
Interpreting $\int d^nx_c \sqrt{g(x_c)} \sig(x_c)\xpv{\exp(-\frac{1}{\hbar}
S_{int})}_{x_c}$ as a regularization scheme for $\tr\,\sig(\hat{x})$, our 
result shows that different regularization schemes (for example the QFT 
approach and the SI approach) lead to different local trace anomalies. 
Since trace anomalies are not topological, this result could have been 
expected. 

The integrated or global anomaly, for which  $\sig(\hat{x})$ is constant, 
is proportional to the partition function. The answers for this case are 
the same, as they should, but a second subtlety was discovered in the proces. 
Namely, the QFT approach 
has the additional benefit 
that the
transition element $\bra{x_0} \exp(-\frac{\bet}{\hbar}\hat{H}) \ket{x_0}$ 
is 
itself general covariant and one may use normal
coordinates in evaluating Feynman diagrams. The
string-inspired approach uses the cyclic symmetry of the
partition function, which causes many diagrams to vanish
trivially but the integrand $\xpv{ \exp(-\frac{1}{\hbar}S_{int}) }_{x_c}$ 
is not covariant and
the benefits of cyclicity are offset by the fact that one is
not allowed to use normal coordinates.

Our final conlusions are twofold. First, one cannot use the string-inspired 
path integral to construct regulators for local trace anomalies which keep 
Einstein invariance at the quantum level. Second, one can use the 
string-inspired method to evaluate partition functions, but one cannot 
use normal coordinates and the integrands differ from those in the QFT 
approach by total derivatives.

\subsection*{Acknowledgements}

The authors thank the organizers of the 1998 summer workshop on String 
Theory and Black Holes  at
Amsterdam
for their hospitality and discussions.

\end{document}